\documentclass[preprint,aps,nofootinbib]{revtex4-2}
\usepackage{subcaption}
\usepackage[labelformat=parens,labelsep=quad,skip=3pt]{caption}
\usepackage{graphicx}
\usepackage{xcolor}
\usepackage{epstopdf}    % nado dobavit, inache pdf ne budet
\usepackage{dcolumn}% Align table columns on decimal point
\usepackage{amsfonts}
\usepackage{amsmath}
\usepackage{amssymb}
\usepackage{times}
\usepackage{bm}% bold math
\usepackage{color}
\usepackage{comment}

\usepackage{epsf}
\usepackage{float}
\usepackage{enumitem}
\usepackage{xcolor}

\newcolumntype{L}{>{\centering\arraybackslash}m{3cm}}
	
	\newcommand{\bqa}{\begin{eqnarray}}
	\newcommand{\eqa}{\end{eqnarray}}
	\newcommand{\bwt}{\begin{widetext}}
	\newcommand{\ewt}{\end{widetext}}

	\newcommand{\e}{\epsilon}

\newcommand{\edc}{\end{document}}
\newcommand{\bb} {\begin{thebibliography}}
\newcommand{\eb}{\end{thebibliography}}

\newcommand{\bc}{\begin{center}}
\newcommand{\ec}{\end{center}}

\newcommand{\be}{\begin{equation}}
\newcommand{\ee}{\end{equation}\normalsize}
\newcommand{\bea}{\begin{eqnarray}}
\newcommand{\eea}{\end{eqnarray}}
\newcommand{\ba}{\begin{array}{l}   }

\newcommand{\ea}{\end{array}}

\newcommand{\ds} {\displaystyle}
\newcommand{\summa}{\ds\sum}

\def\bfr{{\bf r}}
\def\bfk{{\bf k}}

\newcommand{{\vergul}}{  ,}

\begin{document}

\title{The transition from Bose-Einstein condensate to supersolid states in Rydberg-dressed gases beyond Bogoliubov approximation}
% with weak anisotropies}
\author{Asliddin Khudoyberdiev$^{a}$}\email{asliddinkh@gmail.com}
\author{Zabardast Narzikulov$^{a}$}\email{zabardastn@gmail.com}
\author{B. Tanatar$^{b}$}\email{tanatar@fen.bilkent.edu.tr}

\affiliation{$^a$Institute of Nuclear Physics, Tashkent 100214, Uzbekistan\\
$^b$Department of Physics, Bilkent University,
Bilkent 06800, Ankara, Turkey}

\date{\today}

\begin{abstract}
In this paper, we study Bose-Einstein condensation of Rydberg-dressed atoms considering finite range interactions. We use Hartree-Fock-Bogoliubov approximation based on Mean-Field approach. Moreover, within this approximation  modified by the finite-range character of the two-body interaction  we shall obtain analytical expressions for thermodynamic quantities of Rydberg-dressed Bose gas. The imaginary part of the quasiparticle spectrum of a BEC signals the instability of the roton mode with respect to the formation of supersolid state. Our theory predicts a second-order quantum  phase transition from BEC to supersolid phase for Rydberg-dressed bosons in three dimensions.

\end{abstract}
		\maketitle
%\begin{document}
	
\section{Introduction}
Rydberg atoms are consisting of atoms with a highly excited
electrons. Rydberg atoms are expected to become important tools for quantum information because the manipulation of the entanglement of two or more atoms in these systems are very feasible. Strong and long-range interactions are found in gases of ultracold Rydberg atoms \cite{lesan}. One can assume that they possess interactions via long-range van der Waals (vdW) forces. The inter-atomic interactions of them are much stronger than for the atoms in the ground state. For this reason, theoretical description of these interactions with the behavior of a Bose-Einstein condensate are more complicated than that of locally interacting atoms. In realm of low-temperature physics, ultracold quantum gases realize the order extreme limit for which the interparticle interactions and correlations are typically weak, meaning that classically their range of action is much smaller than the mean interparticle distance. Because of this diluteness, roton excitations are absent in ordinary quantum gases, that is, in Bose-Einstein condensates with contact interactions \cite{chomaz}. Additionally, one usually uses only the Bogolubov approximation for these gases, which is applicable at close to zero temperature and asymptotically weak interactions.

However, about 20 years ago, seminal theoretical works predicted the existence of a roton minimum both in BECs with magnetic dipole-dipole interactions \cite{santos} and in BECs irradiated by off-resonant laser light \cite{odell}. Moreover, a decade ago, a roton softening has been also observed in BECs coupled to an optical cavity \cite{mottl}. It was concluded that the roton spectrum is a genuine consequence of the underlying interactions among particles in dipolar BECs and its minimum shows existence of weakly interacting regime \cite{chomaz}.

In particular, Rydberg atoms also showed to support a roton and maxon modes in their Bogoliubov spectrum \cite{henkel,cormack}. Furthermore, these atoms have been proposed to realize a number of  interesting phases in ultracold gases, such as the supersolid phase \cite{cinti1, cinti2},  metallic quantum solid phase \cite{Li},  and roton excitations \cite{tanatar}. The main difference between Rydberg atoms and dipolar gases  is that the former induce effective, nonlocal interactions, which, opposed to latter with dipolar interactions, are isotropically repulsive \cite{henkel}. Yet, one finds partial attraction in momentum space, giving rise to a roton-maxon excitation spectrum and a transition to a supersolid state in three-dimentional condensates.  Unfortunately, in the case of Rydberg gases, short lifetimes of excited atoms  would be an obstacle in experiments to analyze the spectrum of elementary excitations. A solution to this problem is to weakly dress the ground state with a small fraction of the Rydberg state, which results in several orders of magnitude enhancement of the lifetime  \cite{henkel,tanatar,cormack,pfau}. That is why these atoms are called as $Rydberg$-$dressed$. These timescales enable BEC dynamics with long-range interactions, which is predicted to give rise to phase of novel exotic many-body physics which are mentioned above.

Nevertheless, thermodynamic properties of Rydberg-dressed Bose gases, such as critical temperature, heat capacity, etc., have not been studied yet.  Additionally, the existence of supersolid sate and the phase transition from Rydberg-dressed BECs to supersolid have not been anlyzed in detail. In such a supersolid, the particles that must supply the rigidity to form a crystal, at the same time provide for superfluid nonviscous flow \cite{chester}.  This apparent contradiction continues to attract theoretical interests as well as experimental attempts to analyze this phase briefly.  The existing theoretical studies within the Bogoliubov framework based on the crucial assumption that the true atom-atom interaction can be replaced by a contact (i.e. zero-range) interaction, which, strictly speaking corresponds to point like atoms. However, it is clear that, this assumption can not be justified in the case of Rydberg atoms with a large size.    Therefore, to study the properties of such a condensate, it is necessary to take into account the finite size effects. That is why, we will include these interactions in this present paper.

In this work, we study thermodynamic properties of three dimensional Rydberg-dressed BECs with long-range interactions at zero temperature regime. These analysis were partly done in Bogoliubov approximation which   is valid for low temperatures
and asymptotically weak interactions \cite{cormack}. We shall extend it with a more general approximation, that would be valid for all temperatures and any interaction strength. For this purpose, we use Hartree-Fock-Bogoliubov (HFB) approximation \cite{yukhfb,yukobsor,yukanals,ouraniz,ouraniz2part1} and thus which goes within the mean-filed theory for a wide range of system parameters, such as different particle densities, different soft-core radius and different interaction strength.

This paper is organized as follows. In Section II, the Hamiltonian of the system and the properties of interaction terms are introduced. The particle densities and the excitation of roton and maxon modes are studied in Section III, using more realistic HFB method. In Section IV, we give our main results and their discussions. Finally, in Section V we present our conclusion our work. The application of Hartree-Fock-Bogoliubov approximation based on mean-filed theory in order to obtain the dispersion relations will be given in Appendix A. For a convenience, we give the calculation of our main equations in Appendix B.

\section{Main equations}
We write the Hamiltonian of a system of Rydberg-dressed bosons as
\begin{equation}
H=\int d\bfr\left[\psi^{\dag}(\bfr)\left(-\frac{\nabla^2}{2m}-\mu\right)\psi(\bfr)\right]+ \frac{1}{2}\int d\bfr d\bfr'\left[\psi^{\dag}(\bfr)\psi^{\dag}(\bfr')U(\bfr-\bfr')\psi(\bfr')\psi(\bfr)\right]\, ,
\label{H}
\end{equation}
where $\psi(\bfr)$ is the bosonic field operator, $\nabla^2/2m$ is the kinetic energy operator, $m$ is the mass
of the boson, $\mu$ is the chemical potential, and $U$ is the two-body interaction potential. Here and below we set $\hbar=1$   $k_B=1$.

The interaction potential is divided into two parts with contact interaction and finite range interactions as follows
\begin{equation} \label{pot}
U(\bfr)=g_0\delta(\bfr)+g_2V(\bfr)\, ,
\end{equation}
where $g_0=4\pi a_s/m$, $g_2=\alpha g_0$ with $\alpha$  being the finite-range interaction parameter and $a_s$ is the s-wave scattering length.  Here, the second term corresponds to the van der Waals interaction
\begin{equation} \label{finr}
V(\bfr)=\frac{C_0}{R_c^6+r^6}\, ,
\end{equation}
 where $r$ is the interatomic distance, $C_0$ is the strength of the dressed interaction potential and $R_c$ is the soft-core radius \cite{cormack}. This potential is almost constant at short interparticle distances ($r\ll R_c$) and it has a van der Waals type behavior at long distances ($|r|\gg R_c$). Moreover, while the Rydberg-dressed interaction based on the above potential is purely repulsive in real space, its Fourier transform has a negative minimum at  a finite wave vector $k_{crit}$ and at a critical value of $\alpha_{crit}$. These values help us to analyze the occurrence of different phases in the system.

We now make a Bogoliubov shift for field operators for the occurrence of  BEC, i.e. breaking the gauge symmetry
\begin{equation} \label{shift}
\psi(\bfr)=\sqrt{\rho_0}+\tilde{\psi}(\bfr)\, ,
\end{equation}
where $\sqrt{\rho_0}$ and $\tilde{\psi}(\bfr)$ are the density of condensed particle and field operator of
non-condensed particles, respectively.
We insert (\ref{shift}) into (\ref{H}) such that Hamiltonian is divided into five parts according to powers of field operator $\tilde{\psi}(\bfr)$.
\begin{equation}
H=H_0+H_1+H_2+H_3+H_4\, .  \label{parts}
\end{equation}
Due to the orthogonality principle of the condensate function and the field operator of non-condensed atoms, i.e. quantum conservation condition, Hamiltonians with first and third powers of field operators vanish \cite{yukobsor}. Hence $H_1=H_3=0$ and
\begin{subequations}
\begin{align}
H_0&=\left[-\mu_0\rho_0+\frac{g_0\rho_0^2}{2}\right]\, ,  \label{H0} \\ \nonumber
H_2&=\int d\bfr\left[\tilde{\psi}^{\dag}(\bfr)\left(-\frac{\nabla^2}{2m}-\mu\right)\tilde{\psi}(\bfr)\right]+\frac{\rho_0}{2}\int \left[\tilde{\psi}(\bfr)\tilde{\psi}(\bfr')+\tilde{\psi}^\dag(\bfr)\tilde{\psi}(\bfr)+\tilde{\psi}^\dag(\bfr)\tilde{\psi}(\bfr')\right. \label{H2} \\
&\left.+\tilde{\psi}^\dag(\bfr')\tilde{\psi}(\bfr)+\tilde{\psi}^\dag(\bfr')\tilde{\psi}(\bfr')+\tilde{\psi}^\dag(\bfr)\tilde{\psi}^\dag(\bfr')\right]U(\bfr-\bfr')d\bfr d\bfr'\, ,  \\
H_4&=\frac{1}{2}\int\tilde{\psi}^\dag(\bfr)\tilde{\psi}^\dag(\bfr')U(\bfr-\bfr')\tilde{\psi}(\bfr)\tilde{\psi}(\bfr')d\bfr d\bfr'\, . \label{H4}
\end{align}
\end{subequations}
Now, assuming that the system is uniform, we take the Fourier transforms of the field operators as
\begin{equation} \label{fourier}
\tilde{\psi}(\bfr)=\frac{1}{\sqrt{V}}\summa_ka_k\e^{i\bf k \bfr}\, ,  \quad
\tilde{\psi}^\dag(\bfr)=\frac{1}{\sqrt{V}}\summa_ka^\dag_k\e^{-i\bf k \bfr}\,
\end{equation}
and similar transformation for the interaction potential (\ref{pot})
\begin{equation} \label{tran}
U(\bfr-\bfr')=\frac{1}{V}\summa_kU_k e^{i\bf k(\bfr- \bfr')} , \quad U_k=\int U(\bfr-\bfr') e^{-i\bf k(\bfr- \bfr')}d\bfr\, .
\end{equation}
In momentum space equation (\ref{pot})  takes following form  \cite{cormack}
\begin{equation} \label{fk}
U_k=g_0+g_2F_k , \quad   F_k=\frac{2\pi^2e^{-kR_c/2}}{3kR_c}\left[e^{-kR_c/2}-2\sin\left({\frac{\pi}{6}-\frac{\sqrt{3}kR_c}{2}}\right)\right]
\end{equation}
Here, if the Fourier transform is well defined, one can use the assumption that $g_0=\lim_{k\to 0}U_k$. The numerical analysis
of (\ref{fk}) shows that it includes regions with attractive and repulsive interactions as well, for small momentum.
Hence, there will occur maxon and roton modes in the spectra of elementary excitations \cite{cormack}.
Additionally, it may cause phase transition from BEC to supersolid state at some critical values of $g_2$, i.e. $\alpha$.
For the convenience of the reader, we move the details of  diagonalization of the Hamiltonian in Appendix A.

Thus, for dispersion relation we obtain the following expression
\begin{eqnarray}  \label{dispf}
E_k=\sqrt{\epsilon_k(\epsilon_k+2\Delta_k)}\, ,
\end{eqnarray}
with
\begin{eqnarray}
\label{deltak}
 \Delta_k=(\rho_0+\sigma)U_{k}\, ,
\end{eqnarray}
where  $\epsilon_k=\frac{k^2}{2m}$. $\rho_0=\rho-\rho_1$ and $\sigma$ are condensed and anomalous densities of bosons, given by
\bea \label{ro}
\rho_{1}=\frac{1}{V}\displaystyle{\sum_{\mathbf{k}}}\left\{\frac{W_{k}
[\varepsilon_{k}+\Delta_k]}{E_{k}}-\frac{1}{2}\right\}
\eea
\bea \label{sigm}
\sigma=-\frac{1}{V}\sum_{\mathbf{k}}\Delta_k\frac{W_{k}}{E_{k}}
\eea
where $W_{k}=\coth(E_k/2)/2=f_{B}(E_{k})+{1}/{2}$,
$f_{B}(x)=1/(e^{\beta x}-1)$ with $\beta=1/T$,  is an inverse of temperature $T$. In this article our analysis are based on only zero temperature case, with $W_k=1/2$.  Thus, (\ref{deltak}) become  an equation with respect to $\Delta_k$. For the convenience of the reader, we give  analytical calculations to solve  this equation   in Appendix B in detail.

The energy spectrum in (\ref{dispf}) has a parabolic shape in the case of contact interaction similar to that of free particles. However, as one switches the finite-range interactions,   $E_k$ will have a roton and maxon forms with local minima and maxima, respectively. The imaginary valius of $E_k$ in (\ref{dispf}), which originate from different values of the  interaction strength ($\alpha$), would signal the instability of the homogeneous BEC towards supersolid phase. In fact, in Ref.\cite{rocuz} it was shown that the occurrence of roton instability can cause a first-order phase transition where the ground state changes from a uniform condensate to a supersolid state.

\section{Dispersion relation and densities of particles}
Further requirement for $E_k$ is that at small momentum $k$ the spectrum should be gapless, and, therefore, the phonon dispersion is linear: $E_k \approx ck + O(k^3)$ where $c$ can be considered as the speed of sound. In our case, from (\ref{dispf})
\begin{equation}\label{sound}
 c = \sqrt{\frac{\Delta_0}{m}}\sqrt{1+\frac{2}{3}\pi^2\alpha}
\end{equation}
where $\alpha=g_2/g_0$ and $\Delta_0=\Delta_k|_{g_2=0}$. One can assure that by neglecting
finite-range interactions with $g_2=0$, the well-known sound velocity can be obtained as
$c=\sqrt{\frac{\Delta_0}{m}}$, i.e. $\Delta_0=mc^2$.

For the convenience of the numerical calculations, we introduced dimensionless variables for energies $\tilde{E}=E_k/(g_0\rho)$, $Z=\Delta_k/(g_0\rho)$ and for the soft core-radius $R=R_ck_0$, where $k_0=(6\pi^2\rho)^{1/3}$. Our results confirm that the presence of interaction strength $\alpha$ changes the dispersion relation considerably as it is seen from big dependence of the excitation energy on the momentum, is linear at small values
of interaction strength and at small momenta, as in the case of a BEC with weak interaction.
%%%%%   FIGURE 1
\begin{figure}[H]
     \centering
     \begin{subfigure}[b]{0.49\textwidth}
         \centering
         \includegraphics[width=\textwidth]{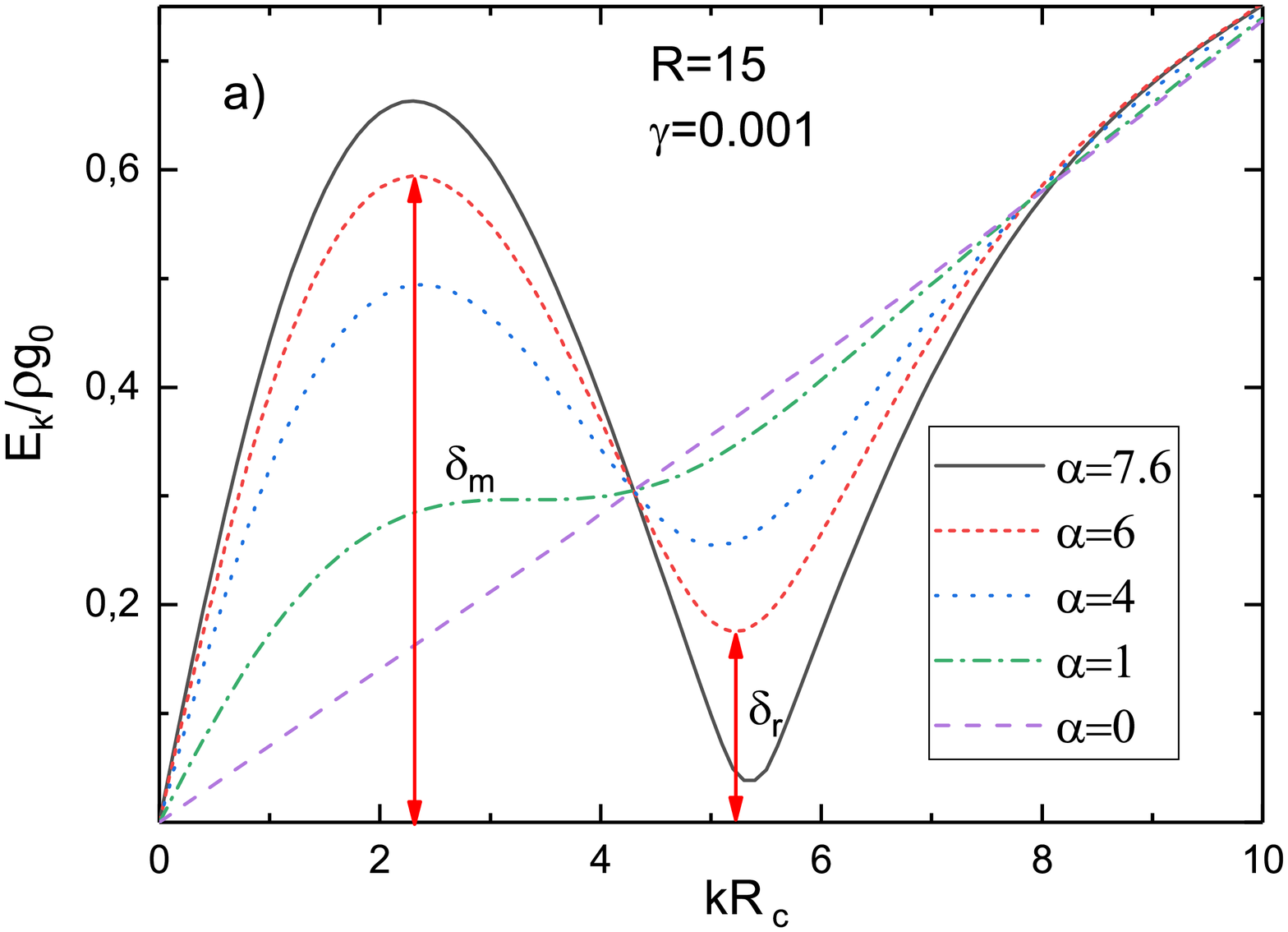}
%         \caption{}
%         \label{fig:alpharotRbsm}
     \end{subfigure}
     \hfill
     \begin{subfigure}[b]{0.49\textwidth}
         \centering
         \includegraphics[width=\textwidth]{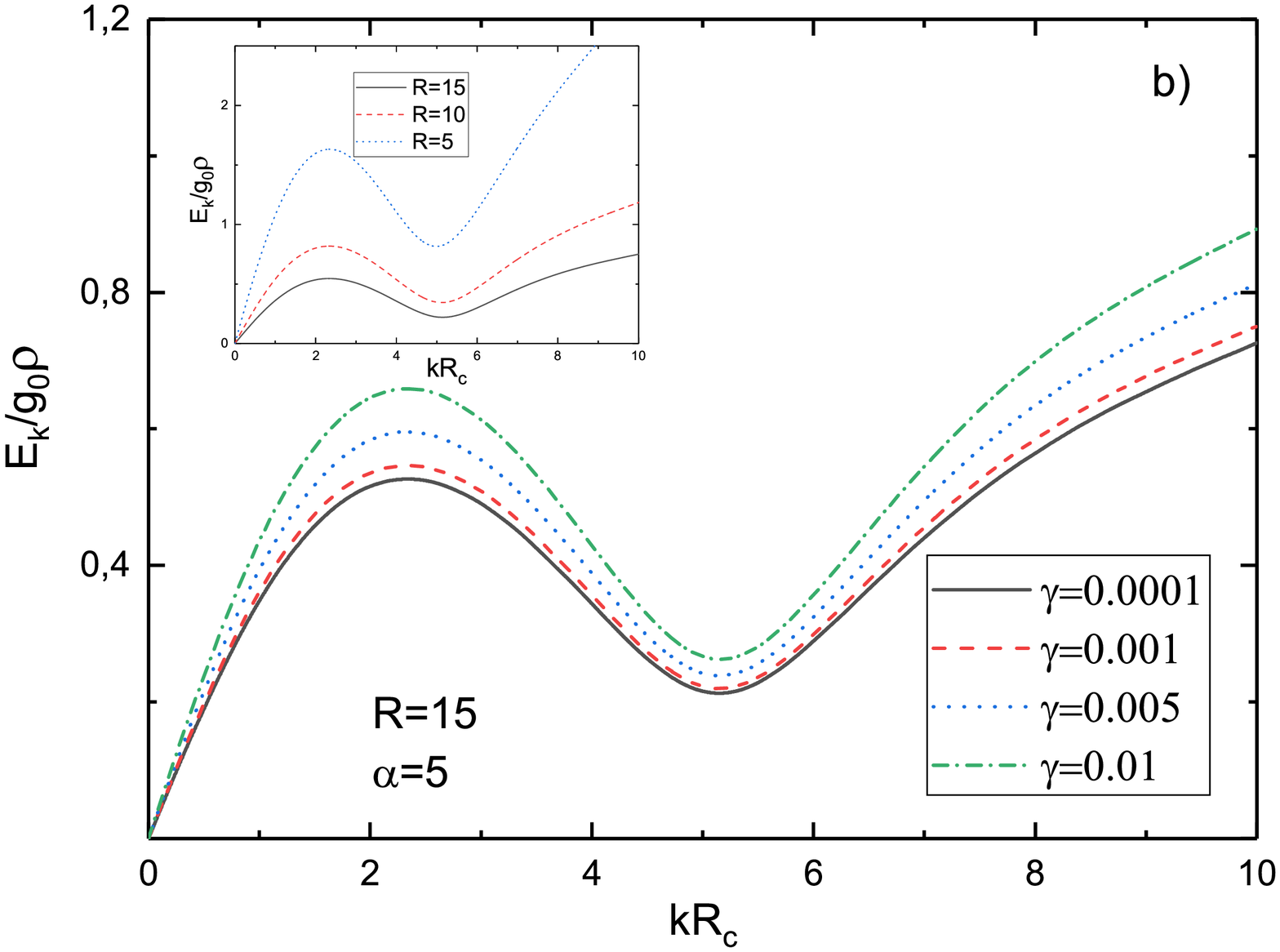}
%         \caption{}
%         \label{fig:alpharotRb}
     \end{subfigure}
     \hfill
        \caption{ Dispersion relation $E_{\bfk}$ (a) for different values of $\alpha$ and for $R=15, \gamma=0.001$
        and (b) for different values of $\gamma$ and  for $R=15, \alpha=5$. Maxon and roton energy gap indicated with arrows. Inset shows the dispersion relation for different values of $R$ and $\gamma=0.001$.}
        \label{fig:disp}
\end{figure}
However, with increasing $\alpha$, there emerge a maximum (maxon mode) and a minimum (roton mode)  in the spectrum. In Fig.\,\ref{fig:disp}a, it can be seen that as $\alpha$ increases, at a certain value ($\alpha_{rot}$),
the spectrum starts to oscillate and the roton and maxon modes appear. The excitation energies at the points of maximum and minimum are noted as the energy of the maxon - $\delta_m$ and roton - $\delta_r$ modes, respectively.  In Fig.\ref{fig:disp}b  the influence of gas parameter to dispersion relation is also considered. It can be seen that only large values of $\gamma$ significantly increase the energies. Additionally, in contrast to gas parameter,  only small
$R$ changes the dispersion relation as shown in the inset.

\begin{figure}[H]
     \centering
     \begin{subfigure}[b]{0.49\textwidth}
         \centering
         \includegraphics[width=\textwidth]{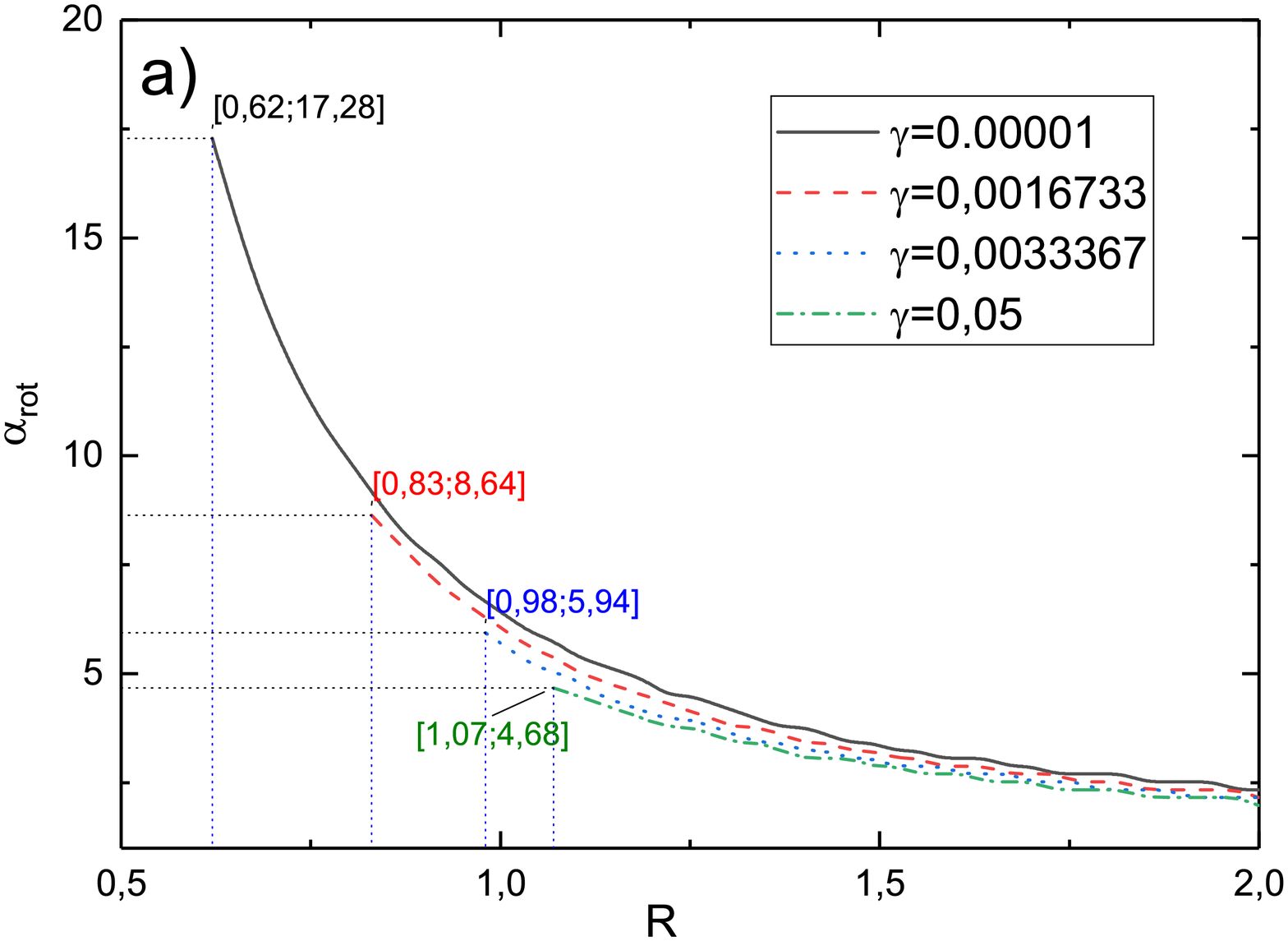}
%         \caption{}
%         \label{fig:alpharotRbsm}
     \end{subfigure}
     \hfill
     \begin{subfigure}[b]{0.49\textwidth}
         \centering
         \includegraphics[width=\textwidth]{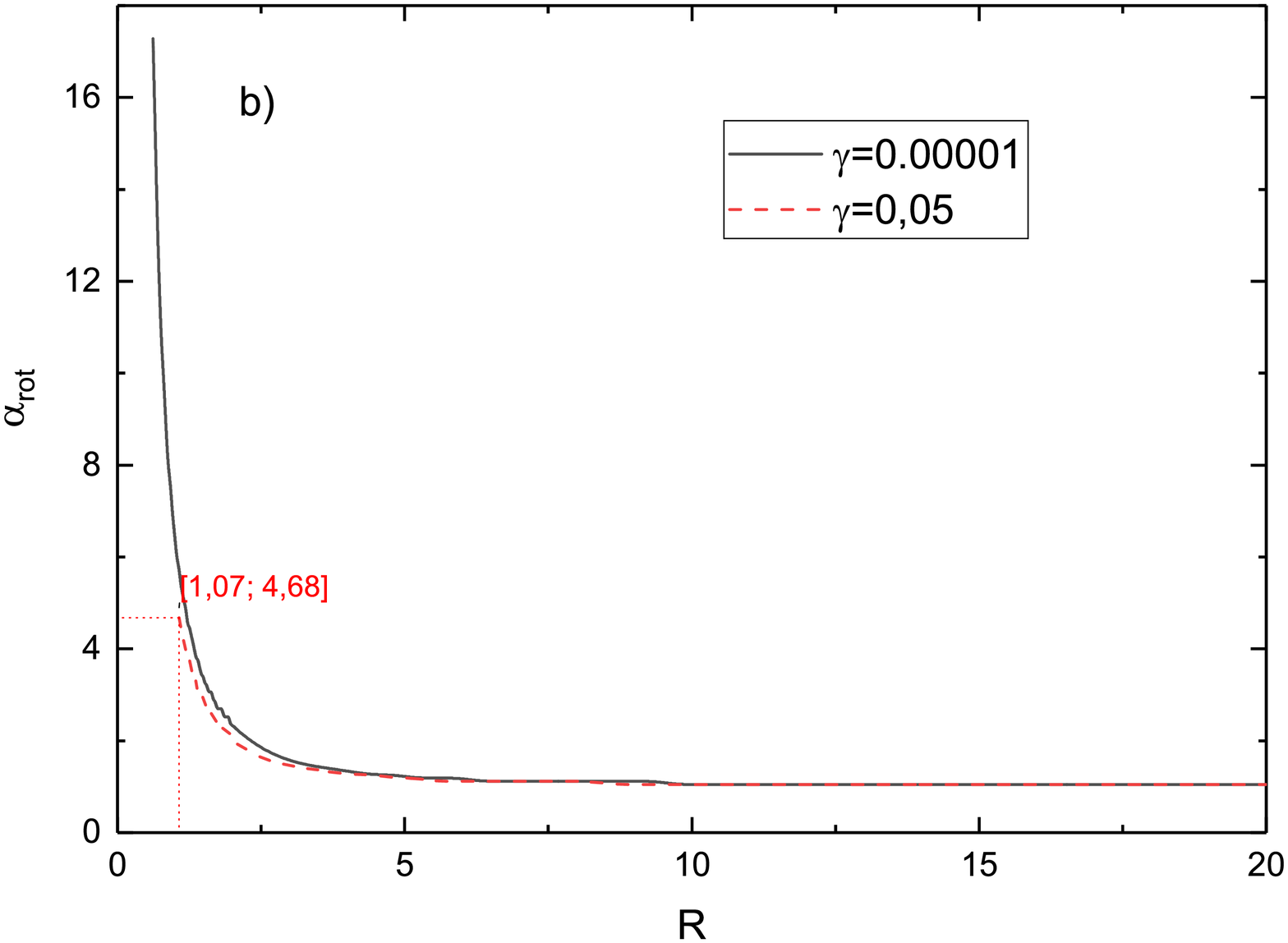}
%         \caption{}
%         \label{fig:alpharotRb}
     \end{subfigure}
     \hfill
        \caption{ $\alpha_{rot}$ vs. $R$ presented for various gas parameter for (a) small and (b) large values of soft core radius. In square brackets shown minimum value of $R$ and corresponding $\alpha_{rot}$ in which roton mode starts for certain $\gamma$. }
        \label{fig:alpharot}
\end{figure}

The value of $\alpha_{rot}$  depends on the soft-core radius $R$ and gas parameter $\gamma$.  In
Fig.\,\ref{fig:alpharot} the dependence of $\alpha_{rot}$ on $R$ is presented for various values of the gas parameter.

In Fig.\,\ref{fig:alpharot}a, $\alpha_{rot}$  is shown for small values of $R$. The curves are similar for different
$\gamma$ (see Fig.\,\ref{fig:alpharot}b), but as $\gamma$ increases, the value of  $\alpha_{rot}$  decreases significantly. These  values for various $\gamma$ are shown in square brackets. At large $\gamma$, roton modes appear at larger values of $\alpha_{rot}$  for smaller $R$. In contrast, as it is seen from Fig.\,\ref{fig:alpharot}b that  other values of $R$ and $\alpha$ lead to non-physical results for the density of condensed particles ($\rho_{0}/\rho<0$). Hence, the quantities of these parameters can be considered as critical ones when the quantum fluctuation  $\rho_1/\rho$ reaches its maximum value \cite{ourfluc}. At higher densities of non-condensed particles, excitations with momenta $k$  become unstable and the condensate disappears.

\begin{figure}
        \centering
        \begin{subfigure}[b]{0.475\textwidth}
            \centering
            \includegraphics[width=\textwidth]{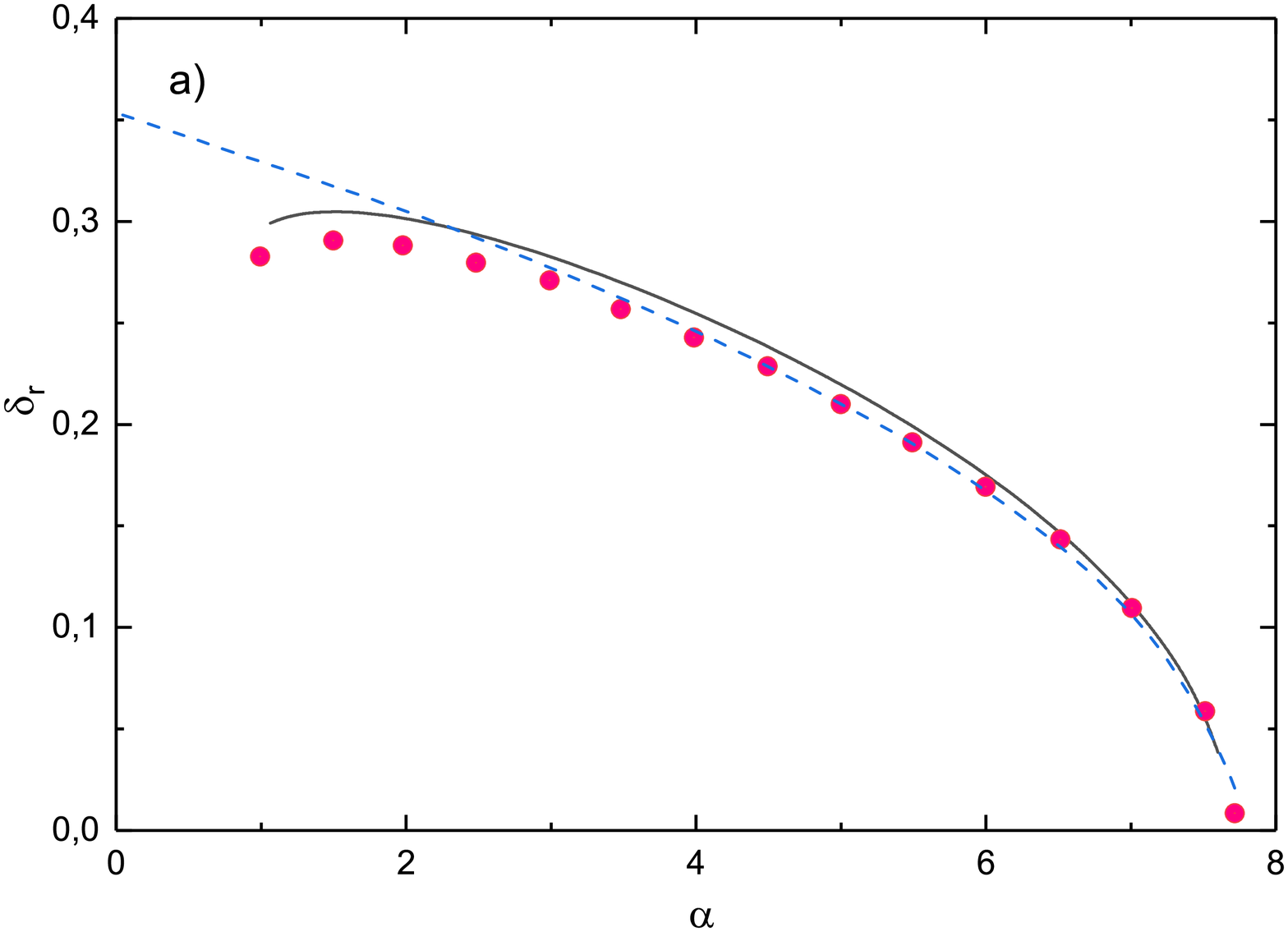}

       \end{subfigure}
        \hfill
        \begin{subfigure}[b]{0.475\textwidth}
            \centering
            \includegraphics[width=\textwidth]{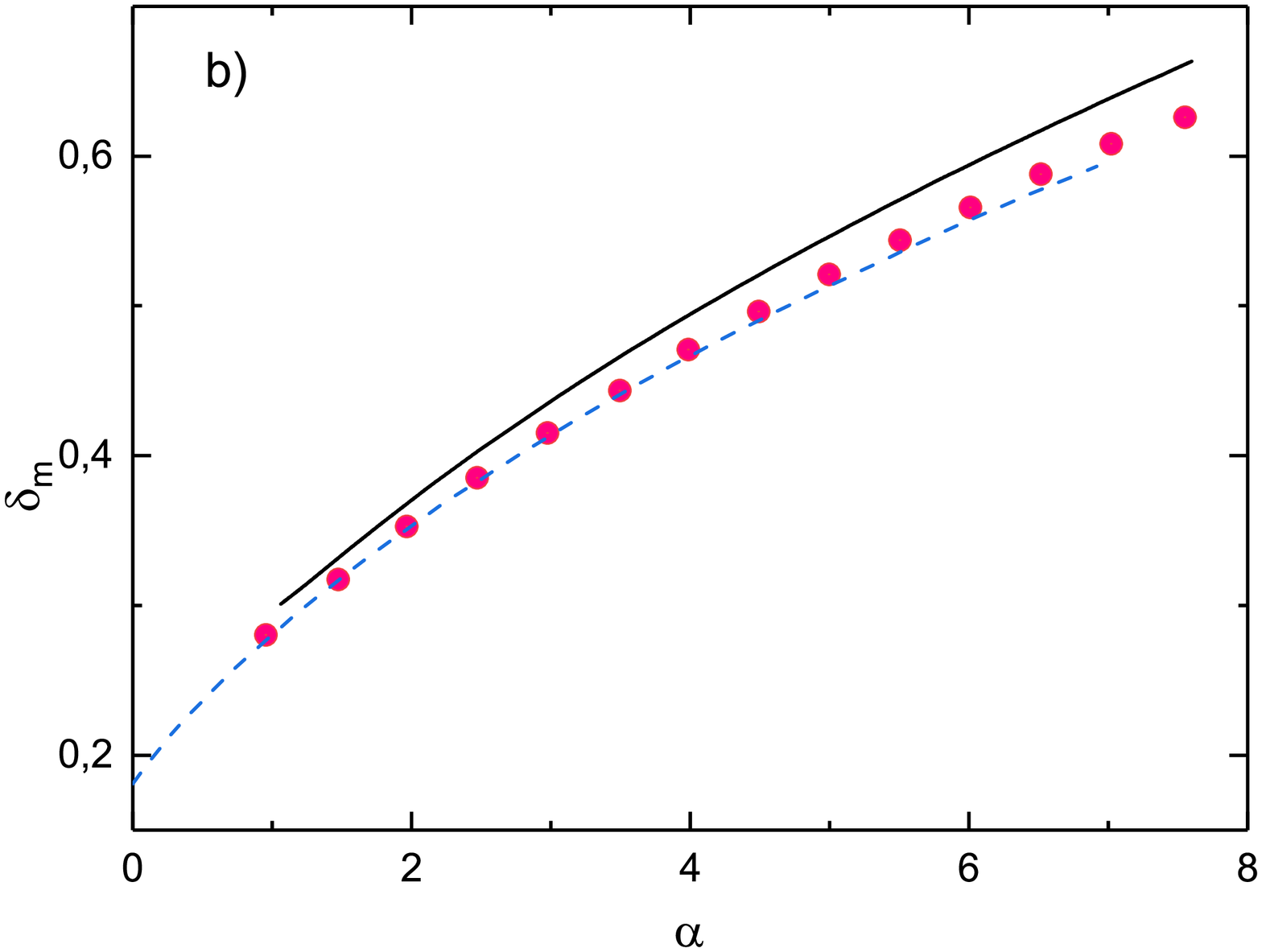}

        \end{subfigure}
        \vskip\baselineskip
        \begin{subfigure}[b]{0.475\textwidth}
            \centering
            \includegraphics[width=\textwidth]{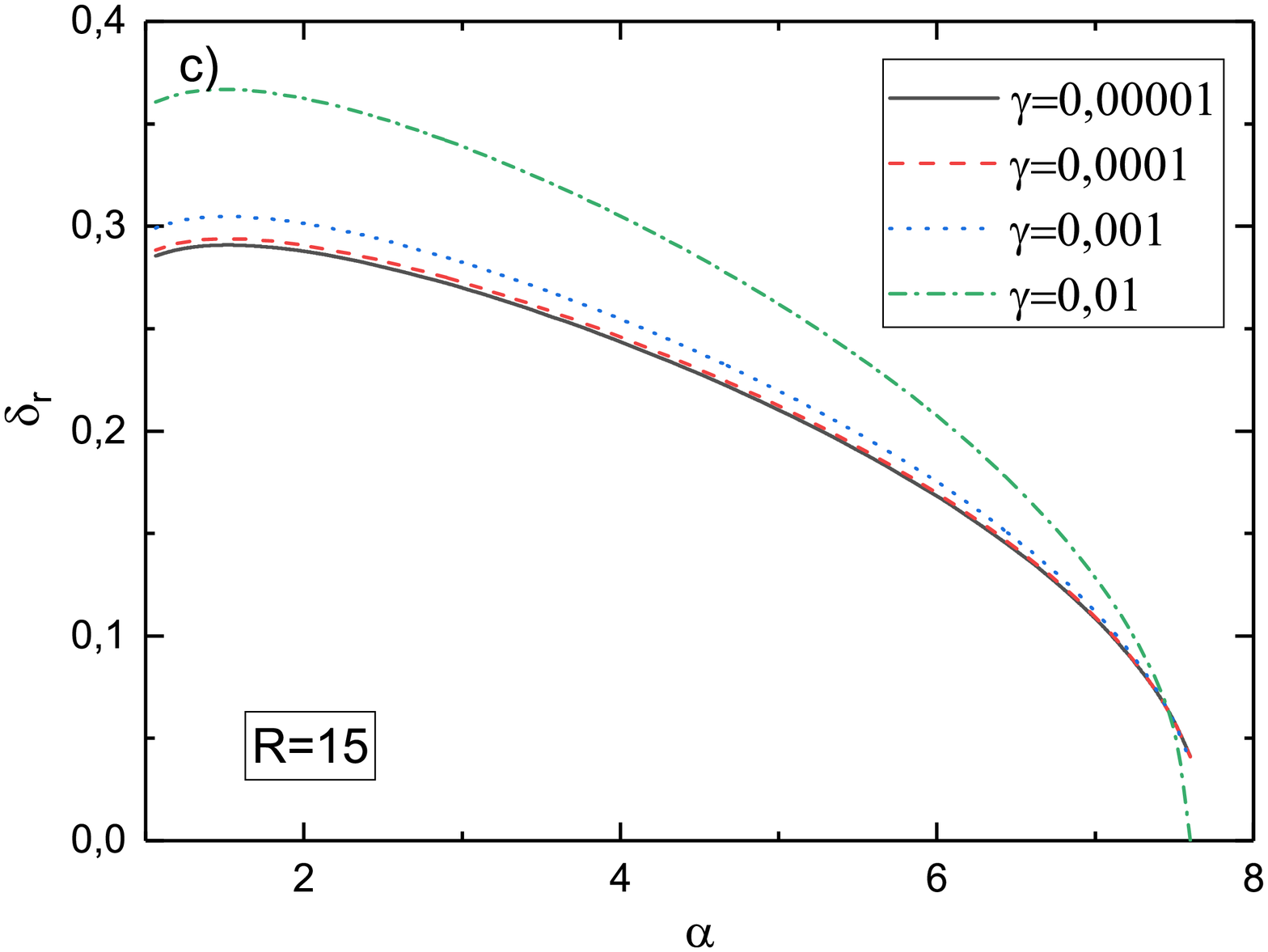}

        \end{subfigure}
        \hfill
        \begin{subfigure}[b]{0.475\textwidth}
            \centering
            \includegraphics[width=\textwidth]{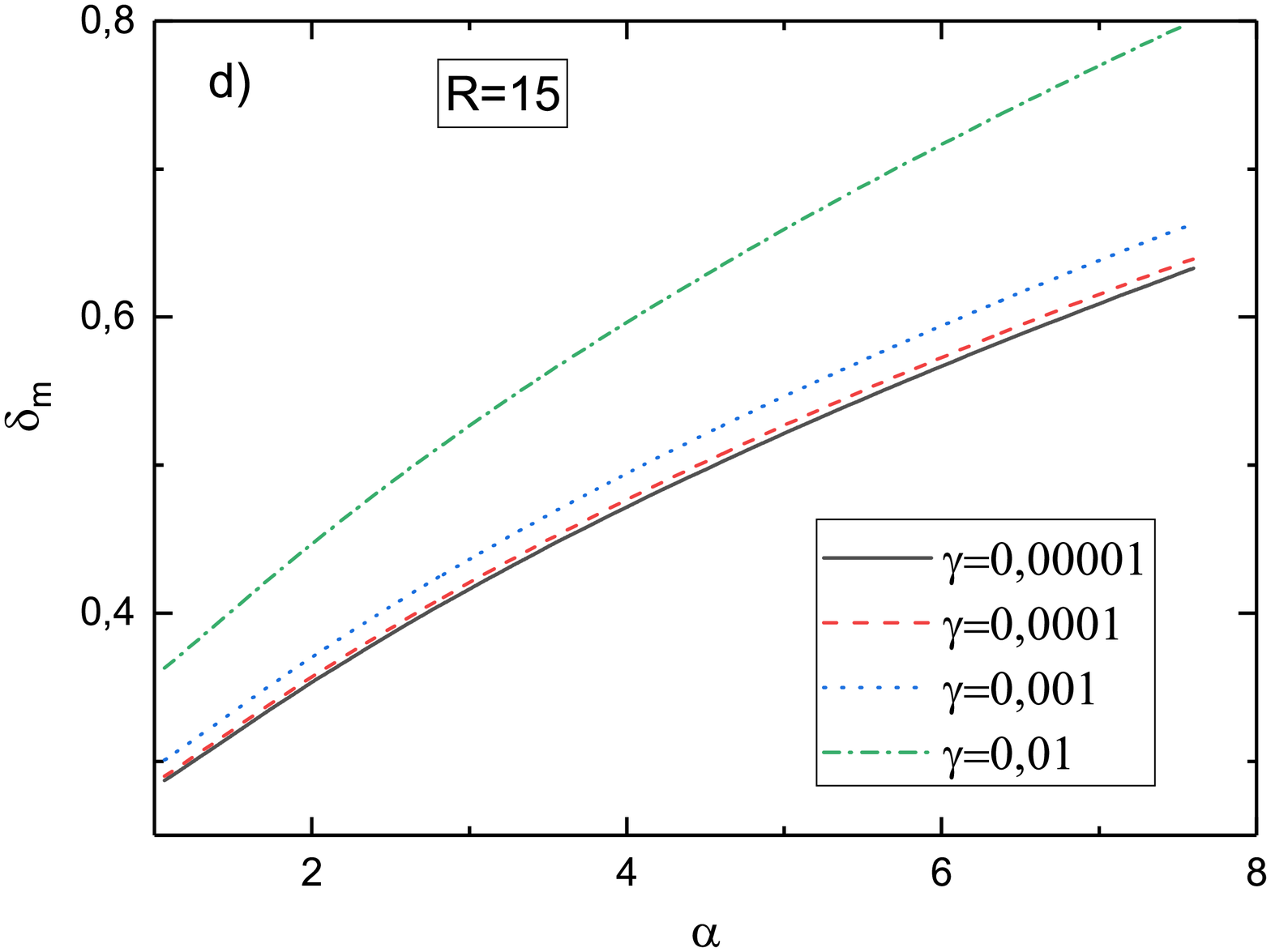}
        \end{subfigure}
        \caption{ Roton and maxon energies. (a) Roton   mode energy depending on $\alpha$ (black solid line) compared with analytical (dashed blue line) and numerical (doted red line) results based on Bogoliubov approach taken from \cite{cormack}. Here $R=15$ and $\gamma=0.001$. (b) The same as in (a), but for maxon mode energy. (c) Roton   mode energy depending on $\alpha$ for different $\gamma$, (d) The same as in (c), but for maxon mode.}
        \label{fig:rotmaxalpha}
    \end{figure}

Roton and maxon energies depend on also interaction strength as it shown in Figs.\ref{fig:rotmaxalpha}.   Our results with HFB approach (black solid lines) are compared with Bogoliubov approach and outcomes are given in Figs.\ref{fig:rotmaxalpha}a,b. Both analytical calculations (blue dashed lines) and  numerical data (red dots) are  taken from \cite{cormack}. The numerical data of Bogoliubov approach for the roton  mode (Fig.\ref{fig:rotmaxalpha}a ) are in agreement completely with our results. The emergence of roton/maxon modes starts from the certain $\alpha$ and $R$ (see. Fig.\ref{fig:alpharot}). This can explain the deviation of analytical calculations in the Bogoliubov approximation, since the curve does not start from $\alpha=0$. Meanwhile, for maxon energy, the similar constraints on small $\alpha$ are also justified (Fig.\ref{fig:rotmaxalpha}a). Furthermore, HFB and Bogoliubov  approaches give  quite similar results  as it is shown in Fig.\ref{fig:rotmaxalpha}b.

In Figs.\ref{fig:rotmaxalpha}c,d,  the energies of roton and maxon with dependence on $\alpha$  are presented at various $\gamma$. One can  see that the deviation occurs only at large $\gamma$. Hence, it is difficult to vary the energies of roton and maxon by tuning the scattering length in $\rho= \gamma/a_s^3$. From Figs.\ref{fig:rotmaxalpha}a,c it is seen that with increasing $\alpha$, the roton energy  $\delta_r$ decreases. For
sufficiently larger $\alpha$, the roton gap   vanishes since the excitation energy becomes complex.  McCormack et.al. \cite{cormack} proposed the way to obtain the critical value of $\alpha=\alpha_{inst}$, where  roton mode become  unstable. According to their method, the Fourier transform of the soft-core potential has the most negative value around $k_r\approx 5\pi/3R_c$ and the roton minimum takes place around this momentum.  Our HFB approach gives the following result for the  $\alpha_{crit}$

 \bea
 \alpha_{crit}=\frac{5e^{5\pi/3}
  \left(36R^2 \tilde{Z}+25\pi^2\right)}
  {72\pi R^2\tilde{Z}
  \left[2e^{5\pi/6}\sin{\left(\frac{\pi}{6}-\frac{5\pi}{2\sqrt{3}}\right)}-1\right]}.
   \eea
or, more simple form
 \bea
 \alpha_{crit}=7.498+\frac{51.39}{R^2\tilde{Z}}
  \eea
where  $\tilde{Z}$ is the solution of Eq.\,(\ref{zbareq}).

\begin{figure}[h]
     \centering
    \includegraphics[width=0.6\textwidth]{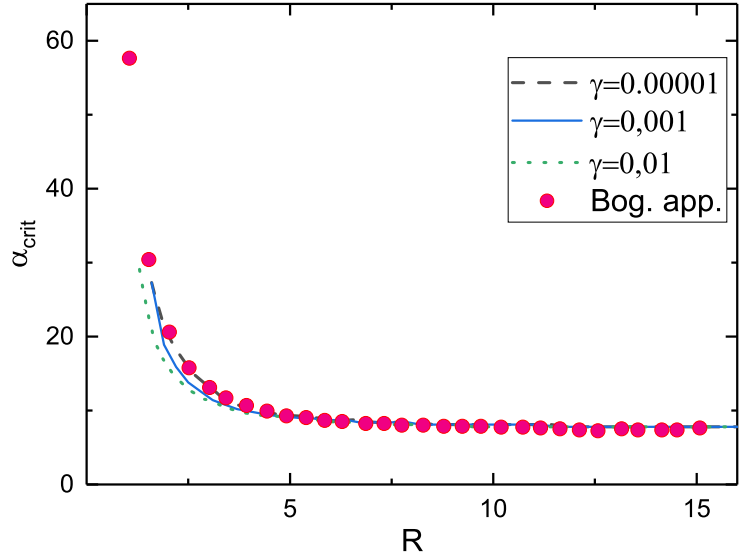}
    \caption{Critical $\alpha$ versus soft core radius for different gas parameter compared numerical data (red doted line) taken form \cite{cormack}.}
    \label{fig:alphacrt}
\end{figure}

In Fig.\,\ref{fig:alphacrt} dependence of critical $\alpha$ on soft-core radius is presented for various values of
$\gamma$. It is seen that for small $R$, interaction strength can reach large values, but for $R \to \infty$,
$\alpha_{crit}\approx 7.5$. Another truth-worthy result of our theory is that $\alpha_{crit}$ vs. $R$ is in good agreement with numerical results of Ref. \cite{cormack} for small $\gamma$. For large $\gamma$  there is a slight deviation in small $R$ region, originating from the HFB approximation.

\begin{figure}
        \centering
        \begin{subfigure}[b]{0.475\textwidth}
            \centering
            \includegraphics[width=\textwidth]{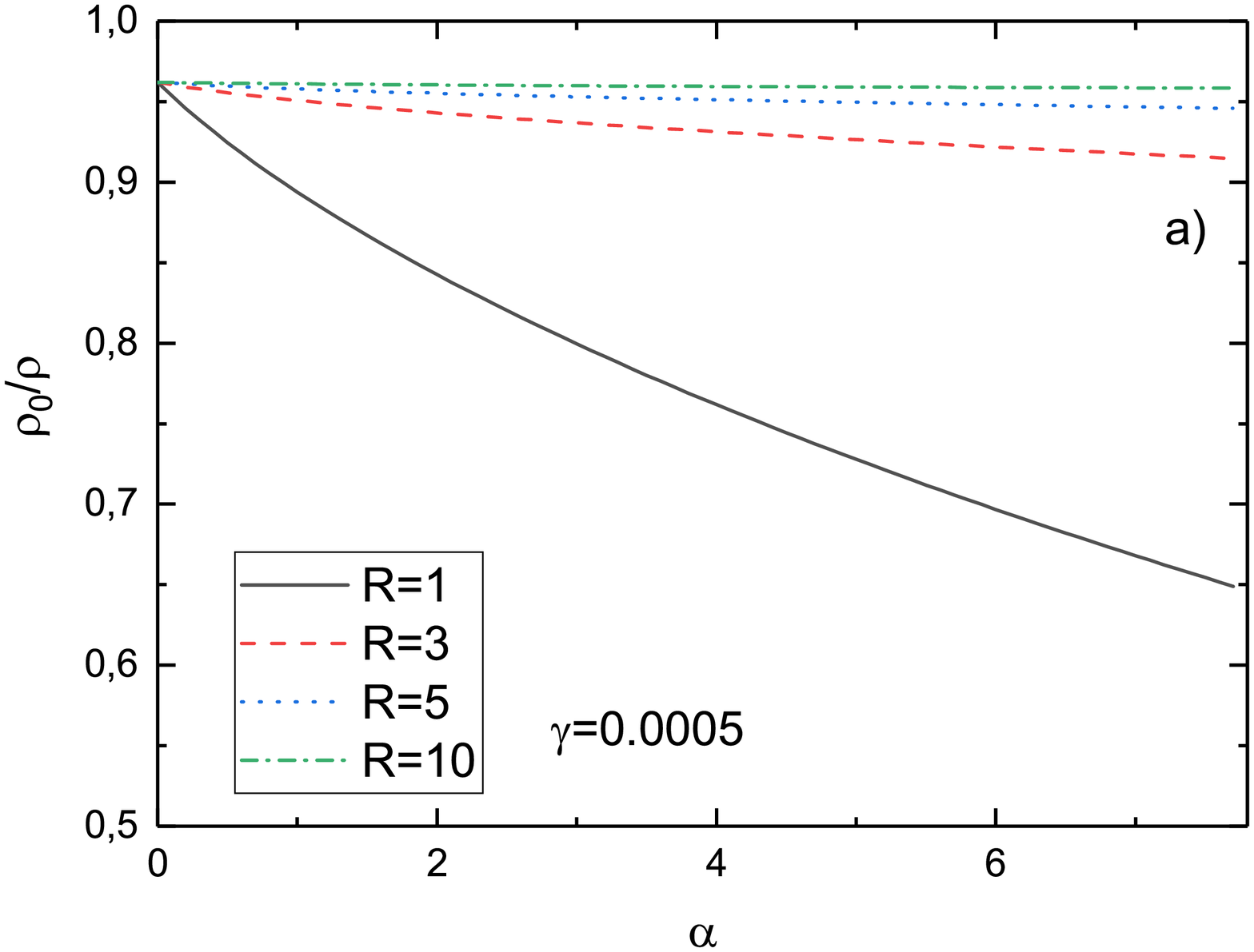}

       \end{subfigure}
        \hfill
        \begin{subfigure}[b]{0.475\textwidth}
            \centering
            \includegraphics[width=\textwidth]{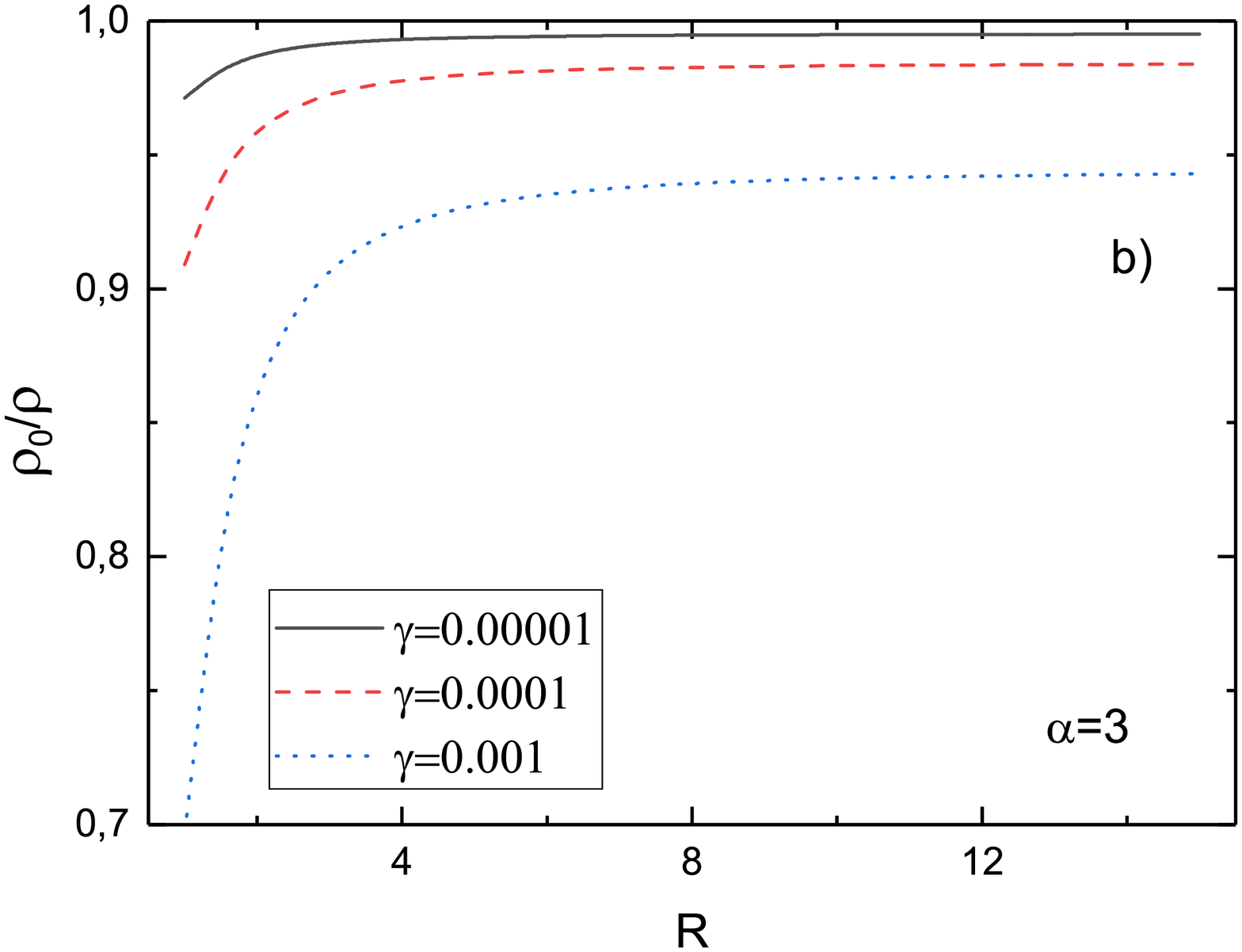}

        \end{subfigure}
        \vskip\baselineskip
        \begin{subfigure}[b]{0.475\textwidth}
            \centering
            \includegraphics[width=\textwidth]{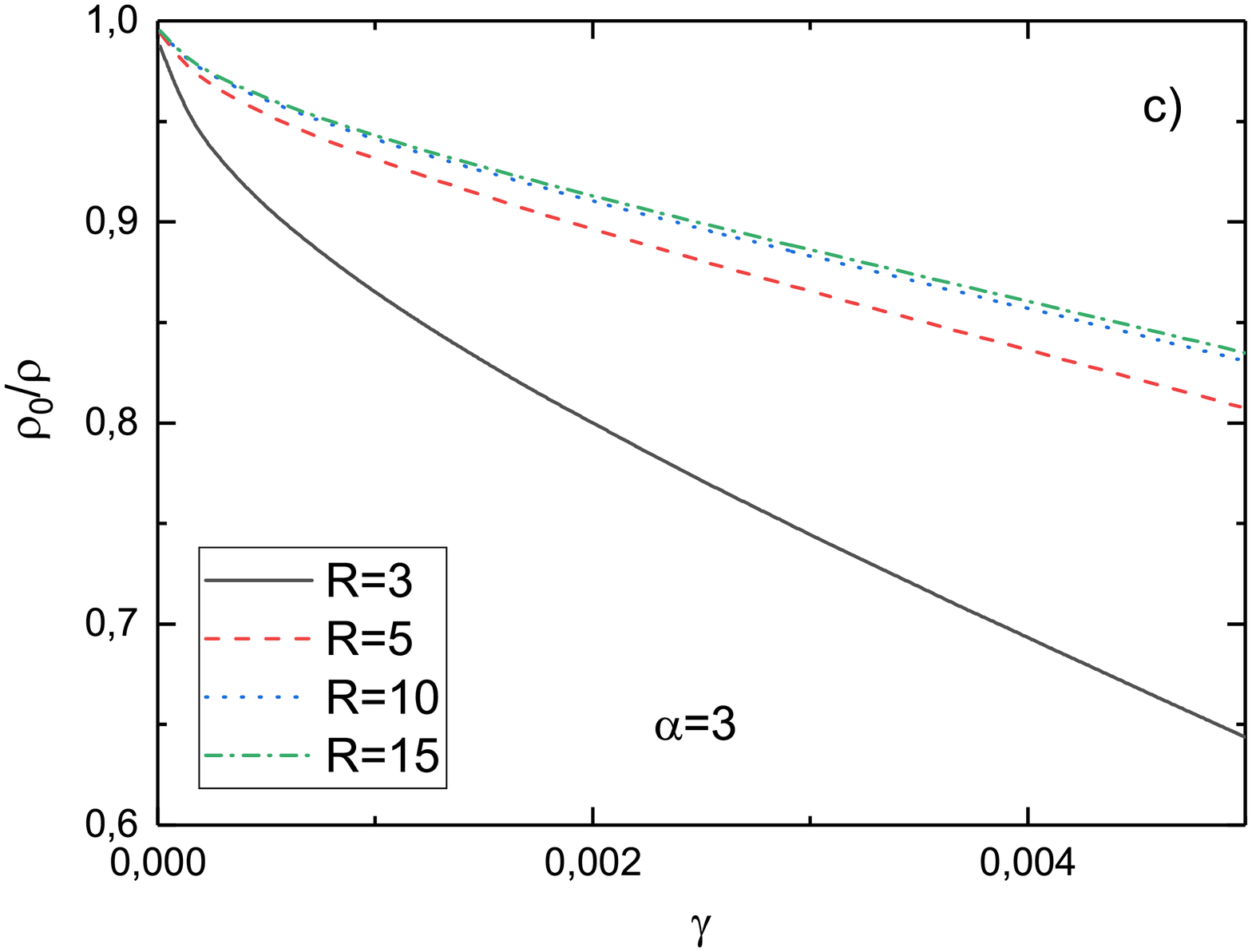}

        \end{subfigure}
        \hfill
        \begin{subfigure}[b]{0.475\textwidth}
            \centering
            \includegraphics[width=\textwidth]{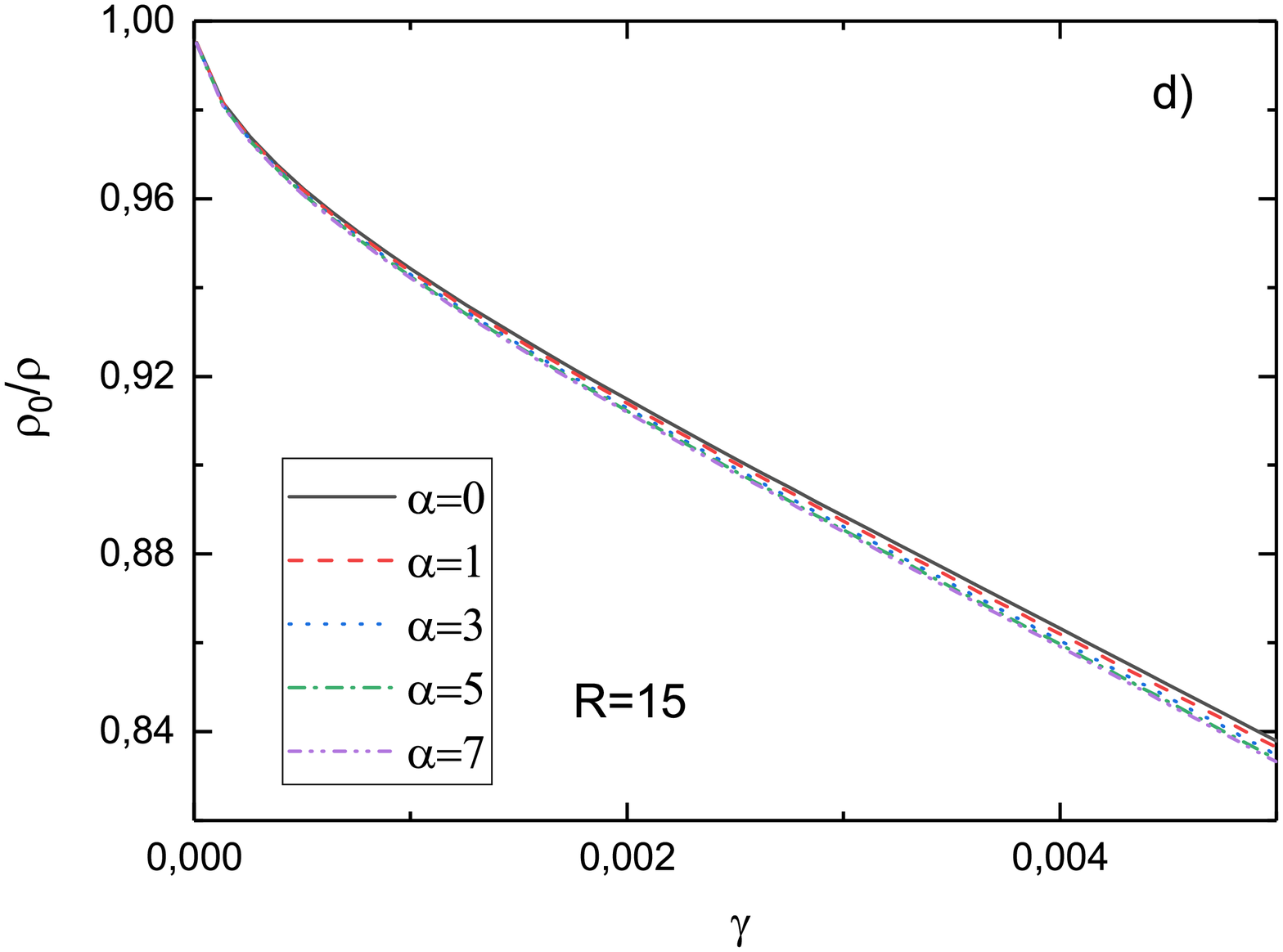}

        \end{subfigure}
        \caption{Density of condensed particles versus various input parameters. (a) $\rho_0/\rho$ vs. $\alpha$ for different $R$ and fixed $\gamma$. (b) Density vs. soft core radius for different $\gamma$. (c) Density as a function of gas parameter for different $R$. (d) Same as (c) but for different $\alpha$.}
        {\small }
        \label{fig:rho0}
    \end{figure}

It is interesting to know how interaction strength affects the density of condensed (uncondensed) atoms. In Figs.\ref{fig:rho0} density of condensed atoms depending on various parameters are presented. Our results show that an increase in the interaction strength makes the  condensate fraction to decrease. But this disproportionality is  significant and quite true for the small values of soft-core radius.  From Fig.\ref{fig:rho0}d, it can be seen that  the interaction strength does not  modify considerably the condensate fraction in contrast to small values of soft core radius (see Fig.\ref{fig:rho0}b).   Starting from $R= 8$,  the condensate fraction practically does not change and remain at  its maximum value. However, its value depletes sharply  with a decreasing soft core radius. On the other hand, this depletion is more sensitive to gas parameter (Fig.\ref{fig:rho0}c). Furthermore, we have shown that (see Fig.\ref{fig:rho0}c,d)) even at large values of $\gamma$, there is no instability in  BEC.

\begin{figure}[h]
     \centering
     \begin{subfigure}[b]{0.49\textwidth}
         \centering
         \includegraphics[width=\textwidth]{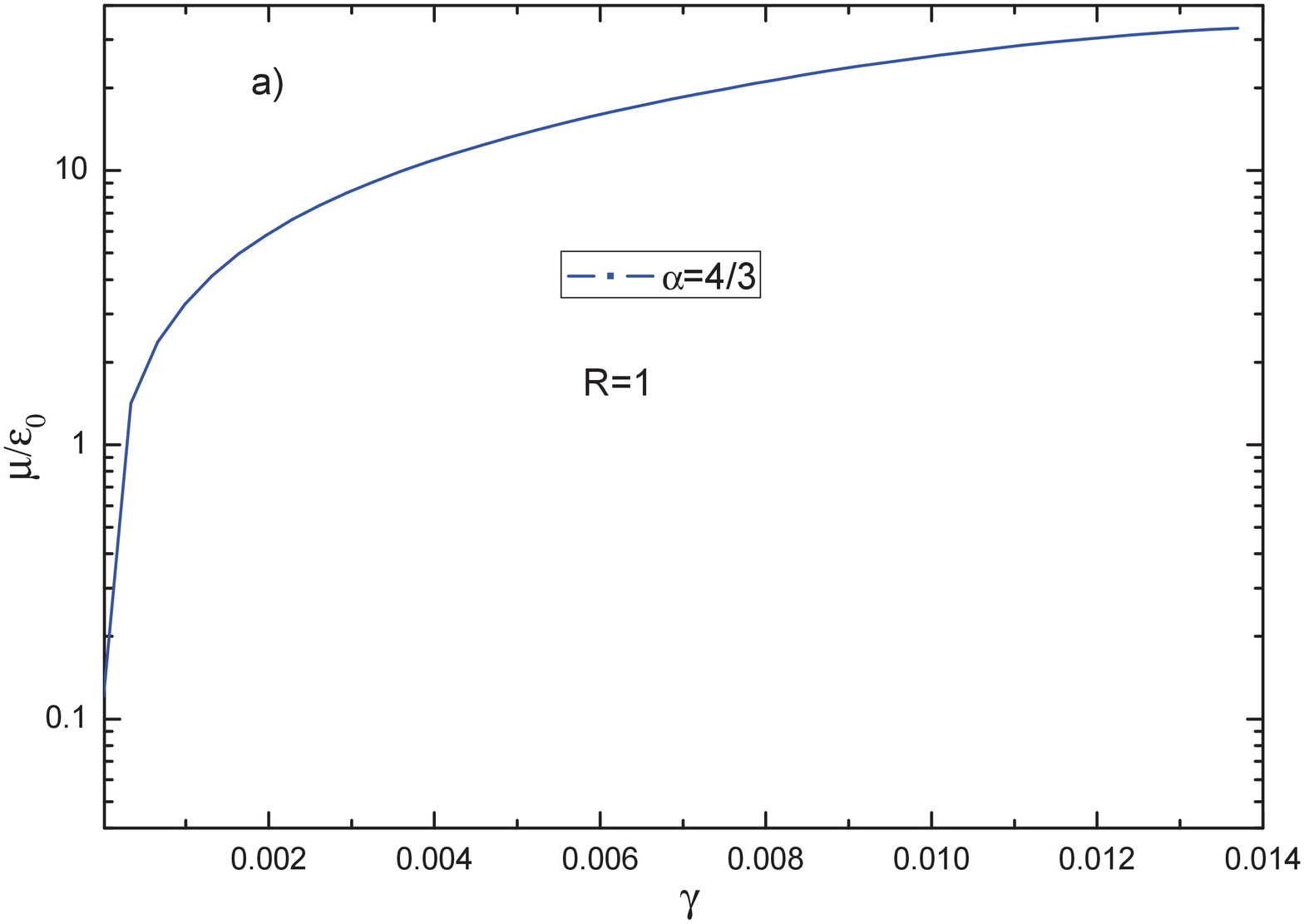}
%         \caption{}
%         \label{fig:alpharotRbsm}
     \end{subfigure}
     \hfill
     \begin{subfigure}[b]{0.49\textwidth}
         \centering
         \includegraphics[width=\textwidth]{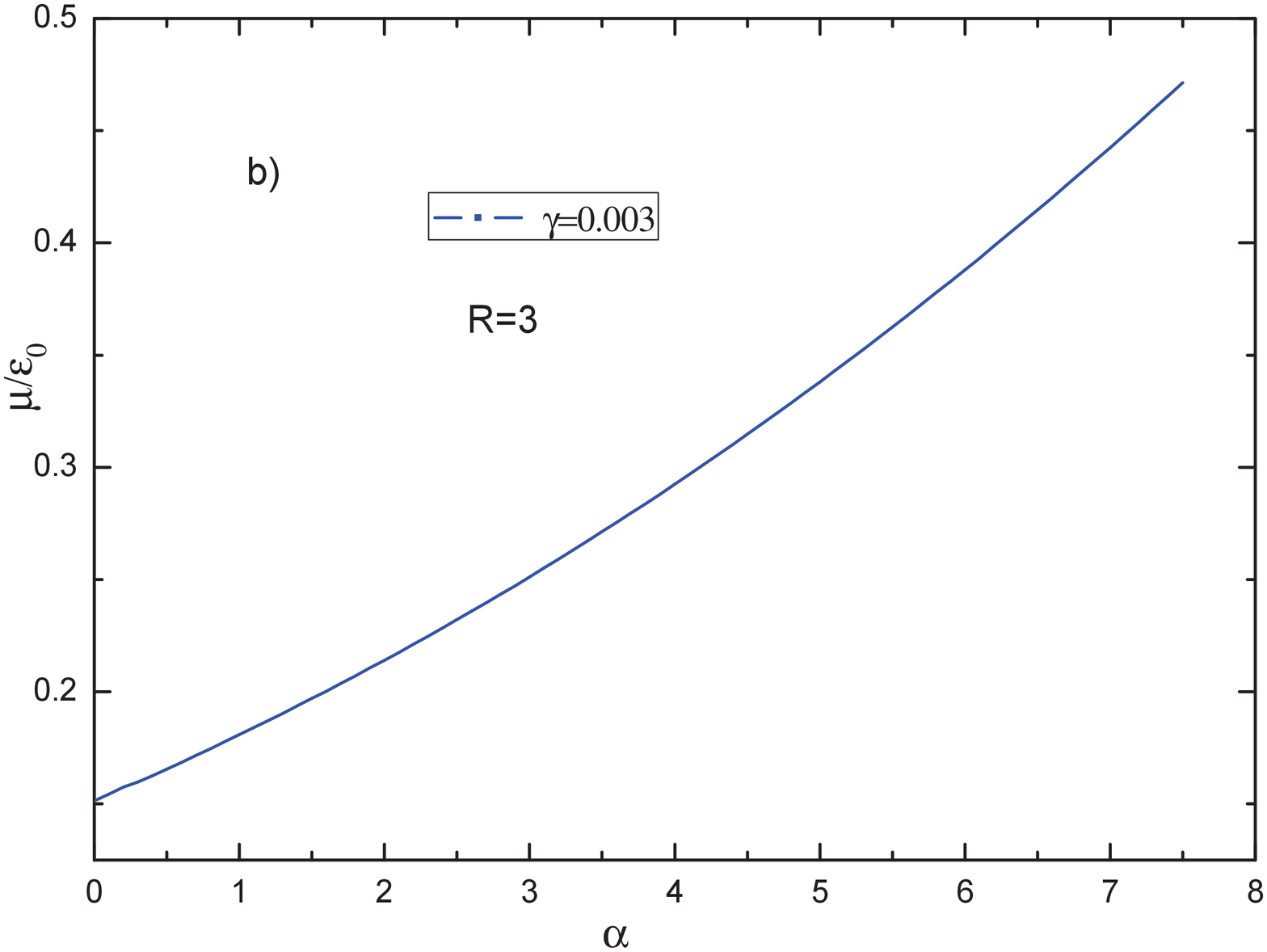}
%         \caption{}
%         \label{fig:alpharotRb}
     \end{subfigure}
     \hfill
        \caption{ Chemical potential dependence on gas parameter $\gamma$ (a) and finite-range interaction strength $\alpha$ (b). }
        \label{fig:chemp}
\end{figure}

For more deeper analysis, we consider here the chemical potential dependence on $\gamma$ and $\alpha$ (see Figs.\ref{fig:chemp}). Our results agree with the results of Quantum Monte Carlo simulations which is performed with a hard-core finite range potential \cite{salash}.

%\newpage
\section{conclusion}
In this work, we proposed a mean-field theory based approach within the Hartree-Fock-Bogoliubov approximation,
which takes into account an anomalous density $\sigma$, for  Rydberg-dressed Bose gases with soft-core
interactions. Our results show that the energy spectra of this system supports both roton and maxon modes
originating from  finite range interactions between atoms. While the maxon energy increases as the the interactions
become strong, the roton energy goes through a minimum. However, for the larger interaction strengths,
this spectrum gives negative results for lower roton modes which signals the occurrence of supersolid phase.
Moreover, we have shown that the condensate fraction can be controlled by s-wave scattering length, i.e. gas parameter $\gamma$. Rather small quantities of the soft core radius also modify the density of condensed particles. However, there is no sensitivity to the strength of the interactions $\alpha$ within the Rydberg-dressed gas.

In a future work, we plan to cover  finite temperature regime for this analysis as well as to  carry on theoretical research for the occurrence of droplets in Rydberg-dressed gases.
%\newpage
\acknowledgments

We express our sincere thanks to Abdulla Rakhimov for his discussions and comments. This work is supported by the Ministry of Innovative Development of the Republic of Uzbekistan and the Scientific and the Technological Research Council of Turkey (TUBITAK)
under Grant No.\,119N689.

\appendix
\numberwithin{equation}{section}
\section{ Diagonalization of the total Hamiltonian}

We rewrite Hamiltonians (\ref{H2}) and (\ref{H4}) using (\ref{fourier}). $H^{(2)}$ and $H^{(4)}$ in second quantization take the form
\begin{eqnarray}
\label{h21}
&H^{(2)}&=\summa_k\left[\frac{k^2}{2m}+\rho_0g_0+\rho_0U_k-\mu\right]a^\dagger_ka_k+\frac{\rho_0}{2}\summa_kU_k(a^\dagger_ka^\dagger_{-k}+a_ka_{-k})\, , \\
&H^{(4)}&=\frac{1}{2V}\summa_q\summa_{k,p}U_ka^\dagger_ka^\dagger_pa_{p+q}a_{k-q}\, . \label{h44}
\end{eqnarray}

To diagonalize $H$ we use the prescription based on the Wick's theorem \cite{ueda}.  The main idea is that the
higher-order term (\ref{h44}) can be made into quadratic forms by
applying the Hartree-Fock Bogoliubov approximation to non-condensed
atoms:
\begin{eqnarray} \nonumber
&a^\dagger_ka^\dagger_pa_{p+q}a_{k-q} \approx n_{k,p+q}a^\dagger_pa_{k-q}\delta_{k,p+q}+n_{p,k-q}a^\dagger_ka_{p+q}\delta_{p,k-q}-n_{k,p+q}n_{p,k-q}\delta_{k,p+q}\delta_{p,k-q}\\ \nonumber
&+n_{k,k-q}a^\dagger_pa_{p+q}\delta_{k,k-q}+n_{p,p+q}a^\dagger_ka_{k-q}\delta_{p,p+q}-n_{k,k-q}n_{p,p+q}\delta_{k,k-q}\delta_{p,p+q}+\sigma_{k,p}a_{p+q}a_{k-q}\delta_{-k,p}\\
&+\sigma_{p+q,k-q}a^\dagger_ka^\dagger_p\delta_{-p-q,k-q}-\sigma_{k,p}\sigma_{p+q,k-q}\delta_{-k,p}\delta_{-p-q,k-q}   \label{wick}
\end{eqnarray}
where $\langle a^\dagger_ka_p\rangle=\delta_{k,p}n_k$, $\langle a_ka_p\rangle=\delta_{-k,p}\sigma_k$  with $n_k$ and $\sigma_k$ are normal ($\rho_1=\frac{1}{V}\summa_kn_k$), and anomalous ($\sigma=\frac{1}{V}\summa_k\sigma_k$)
densities \cite{yukanals,yukobsor,ouraniz,ouraniz2part1, ouraniz2part2,ourmce}. Inserting  (\ref{wick}) into (\ref{h21}) and (\ref{H0}) accordingly, our Hamiltonians with zero-order and second-order terms take the following forms
\begin{eqnarray}
H^{(0)}_{zo}&=\left[-\mu_0\rho_0+\frac{g_0\rho_0^2}{2}-\frac{\rho_1^2g_0}{2}-\frac{1}{2V}\summa_{k,p}U_k(n_{k+p}n_p+\sigma_{k+p}\sigma_p)\right] \\ \nonumber
H^{(2)}_{so}&=\summa_k\left[\frac{k^2}{2m}+\rho g_0+\rho_0U_k+\frac{1}{V}\summa_pn_pU_{k+p}-\mu\right]a^\dagger_ka_k+\\
&+\frac{1}{2}\summa_k\left[\rho_0U_k+\frac{1}{V}\summa_p\sigma_pU_{k+p}\right](a^\dagger_ka^\dagger_{-k}+a_ka_{-k}) \label{Hso} \\ \nonumber
&H_{HFB}=H^{(0)}_{zo}+H^{(2)}_{so}
\end{eqnarray}
where $\rho=\rho_1+\rho_0$ with $\rho$ and $\rho_1$ being the densities of total and non-condensed particles, respectively. To diagonalize the bilinear second-order term ($\ref{Hso}$) we write it in the following form based on HFB approximation
\begin{equation}
H^{(2)}_{so}=\summa_k\omega_ka^\dagger_ka_k+\frac{1}{2}\summa_k\Delta_k(a^\dagger_ka^\dagger_{-k}+a_ka_{-k}) \label{Hso1}
\end{equation}
where
\begin{equation} \label{omega}
\omega_k=\frac{k^2}{2m}+\rho g_0+\rho_0U_k+\frac{1}{V}\summa_pn_pU_{k+p}-\mu\, , \quad  \quad \Delta_k=\rho_0U_k+\frac{1}{V}\summa_p\sigma_pU_{k+p}\, .
\end{equation}
We now introduce the Bogoliubov canonical transformation
\begin{equation}
a_k=u_kb_k+v_kb^\dagger_{-k}, \quad     a^\dagger_k=u_kb^\dagger_k+v_kb_{-k} \label{ankr}
\end{equation}
where the operators $b_k$ and $b^\dagger_k$ can be interpreted as annihilation and creation operators with properties given below
\begin{equation}
[b_k,b^\dagger_p]=\delta_{k,p}\, ,\quad \langle{b^\dagger_kb^\dagger_{-k}\rangle}=\langle{b_kb_{-k}\rangle}=0\, , \quad \langle{b^\dagger_kb_k\rangle}=f_B(E_k)=\frac{1}{e^{E_k/T}-1}\, . \label{requ}
\end{equation}
We insert (\ref{ankr}) into (\ref{Hso1}) and consider (\ref{requ}) as well as the normalization condition $u^2_k-v^2_k=1$. In this case, one can obtain equations for the coefficients $u_k, v_k$
\begin{equation}
\omega_ku_kv_k+\frac{\Delta_k}{2}(u^2_k+v^2_k)=0, \quad u^2_k=\frac{\omega_k+E_k}{2E_k}, \quad v^2_k=\frac{\omega_k-E_k}{2E_k},  \quad
\end{equation}
and these equations yield the dispersion relation, i.e. the spectrum of collective excitations
\begin{equation}
 E_k=\sqrt{\omega^2_k-\Delta^2_k}=\sqrt{(\omega_k+\Delta_k)(\omega_k-\Delta_k)}\, . \label{disp}
\end{equation}
The existence of the Bose-Einstein condensate requires that the spectrum (\ref{disp}) should be gapless. \cite{yukanals}
For this reason, we let $\lim_{k\to 0}E_k=0$, in agreement with Bogoliubov theorem. \cite{bogol} Thus, this condition helps us find an equation for the chemical potential $\mu$, using (\ref{omega})
\begin{equation} \label{mu}
\mu=\rho g_0+\frac{1}{V}\summa_k(n_k-\sigma_k)U_{k}\, .
\end{equation}
In equations (\ref{omega}), one difficulty may arise due to the sum of two wave vectors $\bold k$ and $\bold p$.
Hence, when long-range interactions are considered, the dispersion relation corresponding to the quasiparticle
spectrum of a BEC is qualitatively different, where the excitation energies of the collective modes depend
non-monotonically on the momentum. However, this can be simplified by using the following approximation \cite{yuklaser}
\begin{equation}
\summa_pn_pU_{k+p}\approx U_k\summa_pn_p \quad, \quad  \summa_p\sigma_pU_{k+p}\approx U_k\summa_p\sigma_p
\end{equation}
In this case, our main equations (\ref{omega}) and (\ref{mu}) take the following form
\begin{eqnarray} \label{omega1}
\omega_k=\frac{k^2}{2m}+\rho g_0+\rho U_k-\mu\, , \\  \label{delta1}
 \Delta_k=(\rho_0+\sigma)U_{k}\, .
\end{eqnarray}

\section{ The analytical calculations of our main equation}
Equation for $\Delta_k$, from dispersion relation
\bea \label{delta2}
 \Delta_k=(\rho-\rho_1+\sigma)U_{k}\, .
 \eea
To solve Eq. (\ref{delta2}) with respect to $\Delta_{k}$ we will use following technique: 
first  separate $\rho_1$ and $\sigma$ into two parts ($\bar{\rho}_1$, $\bar{\sigma}$) and ($\rho_{10}$, $\sigma_0$)  as $\rho_1=\bar{\rho}_1+\rho_{10}$ and $\sigma=\bar{\sigma}+\sigma_0$, where zero values correspond to the case without finite range  $g_2=0$. Thus, for $\rho_{10}$ and $\sigma_0$ the summation will easily done by the well-known formulas
\bea \label{rho10}
\rho_{10}=\frac{1}{2V}\displaystyle{\sum_{\mathbf{k}}}\left\{\frac{\varepsilon_{k}+\Delta_0}{E_{k}^0}-1\right\}=\frac{(\Delta_0 m)^{3/2}}{3\pi^2}\, ,
\eea
\bea \label{sigma10}
\sigma_0=-\frac{\Delta_0}{2V}\sum_{\mathbf{k}}\frac{1}{E_{k}^0}=\frac{(\Delta_0 m)^{3/2}}{\pi^2}=3\rho_{10}\, ,
\eea
where $E_k^0=\sqrt{\varepsilon_k(\varepsilon_k+2\Delta_0)}$ and $\Delta_0=g_0(\rho-\rho_{10}+\sigma_0)$.

For ($\bar{\rho}_1$, $\bar{\sigma}$) summation can be done by adding and subtracting zero range  parts from (\ref{ro}) and (\ref{sigm}) and this trick helps to avoid infrared divergence.

The next terms take the form
\bea \label{rho1k}
\bar{\rho}_{1}=\frac{1}{2V}\displaystyle{\sum_{\mathbf{k}}}\left\{\frac{\varepsilon_{k}+\Delta_k}{E_{k}}-\frac{\varepsilon_{k}+\Delta_0}{E_{k}^0}\right\}\, ,
\eea
\bea \label{sigma1k}
\bar{\sigma}=-\frac{1}{2V}\sum_{\mathbf{k}}\left\{\frac{\Delta_k}{E_{k}}-\frac{\Delta_0}{E_{k}^0}\right\}\, ,
\eea
where, $E_k=\sqrt{\varepsilon_k(\varepsilon_k+2\Delta_k)}$  and  $\Delta_k=U_k(\rho-\rho_{1}+\sigma)$.
 The same method is also applied to $\Delta_k$ with and without finite range effect ($\Delta_0$).
Here, we rewrite $\Delta_k$ in the following form
\begin{equation} \label{deltasep}
\Delta_k=\Delta_0(1+\alpha F_k)+g_0(1+\alpha F_k)[\bar{\sigma}-\bar{\rho_1}] =\phi_k+\Phi_k\, ,
\end{equation}
identifying $\phi_k=\Delta_0(1+\alpha F_k)$ and $\Phi_k=g_0(1+\alpha F_k)[\bar{\sigma}-\bar{\rho_1}]$.
It is convenient to introduce a new variable as $I_E=V(\bar{\sigma}-\bar{\rho_1})=\frac{1}{2}\sum_{\mathbf{k}}\frac{E_k^0-E_k}{\varepsilon_k}$.
As a result, we can write that $\Phi_k=g_0(1+\alpha F_k)I_E/V$.

We have also introduced the dimensionless variables $R=R_ck_0$,  $\tilde{E}=E/(g_0\rho)$,   $Z_k=\Delta_k/(g_0\rho)$, $Z_0=\Delta_0/(g_0\rho)$,  where $k_0=(6\pi^2\rho)^{1/3}$, $\rho=\gamma/a_s^3$. Here,  $\gamma$ and $a_s^3$ are the gas parameter and s-wave scattering length, respectively.
Then, Eq.\,(\ref{deltasep}) takes the following form
\bea \label{zk}
Z_k=(Z_0+\frac{I_E}{\rho V})\bar{f}_k\, ,
\eea
where $\bar{f}_k=1+\alpha F_k$. To avoid any dependence on momentum we introduce new variables as $\tilde{Z}=Z_k/\bar{f}_k$ and $\bar{I}_E(\tilde{Z})=\frac{I_E}{\rho V}$. Now, from (\ref{zk}) we obtain
\begin{equation}\label{zbareq}
\tilde{Z}=Z_0+\bar{I}_E(\tilde{Z})
\end{equation}
Firstly, by solving equation for $\Delta_0$ in (\ref{rho10}) one can find $Z_0$. Then,  $Z_k$ i.e. $\Delta_k$ can obtained from Eq.\,(\ref{zbareq}).


\begin{thebibliography}{99}
\bibitem{lesan}I. Lesanovsky, Many-Body Spin Interactions and the Ground State of a Dense Rydberg Lattice Gas, Phys. Rev. Lett. \textbf{106}, 025301 (2011).
\bibitem{chomaz} L. Chomaz, R.M.W. van Bijnen, D. Petter,  G. Faraoni, S. Baier, J. H. Becher, M. J. Mark, F. Wächtler, L. Santos, and F. Ferlaino, Observation of roton mode population in a dipolar quantum gas. Nature Phys. \textbf{14}, 442446 (2018).
\bibitem{santos} L. Santos, G. V. Shlyapnikov, and M. Lewenstein, Roton-maxon spectrum and stability of trapped dipolar Bose-Einstein condensates, Phys. Rev. Lett. \textbf{90}, 250403 (2003).
\bibitem{odell} D. H. J. O'Dell,  S. Giovanazzi, and G. Kurizki,  Rotons in gaseous Bose-Einstein condensates irradiated by a laser, Phys. Rev. Lett. \textbf{90}, 110402 (2003).
\bibitem{mottl}R. Mottl, F. Brennecke, K. Baumann, R. Landig, T. Donner, T. Esslinger, Roton-type mode softening in a quantum gas with cavity-mediated long-range interactions, Science \textbf{336}, 15701573 (2012).
\bibitem{henkel} N. Henkel, R. Nath, and T. Pohl, Three-Dimensional Roton Excitations and Supersolid Formation in Rydberg-Excited Bose-Einstein Condensates,  Phys. Rev. Lett. \textbf{104}, 195302 (2010).
\bibitem{cormack} G. McCormack, R. Nath, and W. Li, Dynamical excitation of maxon and roton modes in a Rydberg-dressed Bose-Einstein condensate, Phys. Rev. A \textbf{102}, 023319 (2020).
\bibitem{cinti1} N. Henkel, F. Cinti, P. Jain, G. Pupillo, and T. Pohl, Supersolid Vortex Crystals in Rydberg-Dressed Bose-Einstein Condensates, Phys. Rev. Lett. \textbf{108}, 265301 (2012).
\bibitem{cinti2} F. Cinti, P. Jain, M. Boninsegni, A. Micheli, P. Zoller, and G. Pupillo, Supersolid Droplet Crystal in a Dipole-Blockaded Gas, Phys. Rev. Lett. \textbf{105}, 135301 (2010).
\bibitem{Li} W.-H. Li, T.-C. Hsieh, C.-Y. Mou, and D.-W. Wang, Emergence of a Metallic Quantum Solid Phase in a Rydberg-Dressed Fermi Gas, Phys. Rev.
Lett. \textbf{117}, 035301 (2016).
\bibitem{tanatar} I. Seydi, S. H. Abedinpour, R. E. Zillich, R. Asgari, and B. Tanatar, Rotons and Bose condensation in Rydberg-dressed Bose gases, Phys. Rev. A \textbf{101}, 013628 (2020).
\bibitem{pfau} J. B. Balewski, A. T. Krupp, A. Gaj, S. Hofferberth, R. Löw and T. Pfau, Rydberg dressing: understanding of collective manybody effects and implications for experiments, New J. Phys. \textbf{16},  063012 (2014).
\bibitem{chester}  A. J. Leggett, Can a Solid Be "Superfluid"? Phys. Rev. Lett. \textbf{25}, 1543 (1970).
\bibitem{yukhfb} V. I. Yukalov and H. Kleinert, Gapless Hartree-Fock-Bogoliubov approximation for Bose gases,
Phys. Rev. A \textbf{73}, 063612 (2006).

\bibitem{yukanals}V. I. Yukalov, Representative statistical ensembles for Bose systems with broken gauge symmetry, Annals of Physics, \textbf{323}, 461-499 (2008).
\bibitem{yukobsor} V. I. Yukalov , Basics of Bose-Einstein Condensation,  Phys. Part. Nuclei \textbf{42}, 460 (2011).
\bibitem{ouraniz} A. Khudoyberdiev, A. Rakhimov, and A. Schilling, Bose-Einstein condensation of triplons with a weakly broken U(1) symmetry, New J. Phys. \textbf{19},  113002 (2017).
\bibitem{ouraniz2part1}  A. Rakhimov, A. Khudoyberdiev, L. Rani, B. Tanatar, Spin-gapped magnets with weak anisotropies I:
Constraints on the phase of the condensate wave
function,  Ann. Phys. \textbf{424},  168361 (2021).


\bibitem{rocuz}S. M. Roccuzzo and F. Ancilotto, Supersolid behavior of a dipolar Bose-Einstein condensate confined in a tube, Phys. Rev. A \textbf{99}, 041601(R)
(2019).

\bibitem{ourfluc} A. Rakhimov,  A. Khudoyberdiev, Z. Narzikulov and B. Tanatar, Defining a critical temperature of a crossover from BEC to the
normal phase in anisotropic quantum magnets, arXiv:2205.13865, (2022).
\bibitem{salash} A. Cappellaro and L. Salasnich, Thermal field theory of bosonic gases with finite-range effective interaction, Phys.Rev.A, \textbf{95}, 033627 (2017).

\bibitem{ueda} N. T. Phuc, Y. Kawaguchi, and M. Ueda, Effects of thermal and quantum fluctuations on the phase diagram of a spin-1 $^{87}$Rb Bose-Einstein condensate
, Phys. Rev. A \textbf{84}, 043645 (2011).
\bibitem{ouraniz2part2} A. Rakhimov,  A. Khudoyberdiev, B. Tanatar, Effects of exchange and weak Dzyaloshinsky-Moriya anisotropies
on thermodynamic characteristics of spin-gapped magnet, Int. J. Mod. Phys. B. \textbf{35}, 2150223 (2021).

\bibitem{ourmce} A. Rakhimov,  A. Gazizulina, Z. Narzikulov, A. Schilling, E. Ya. Sherman, Magnetocaloric effect and Gruneisen parameter of quantum magnets with a spin gap,  Phys. Rev. B \textbf {98} 144416 (2018) .
\bibitem{bogol} N. N. Bogoliubov,  Lectures on Quantum Statistics vol. 2 (New York: Gordon and Breach) (1970).
\bibitem{yuklaser}V. I. Yukalov and E. P. Yukalova, Bose-condensed atomic systems with nonlocal
interaction potentials, Laser Phys. \textbf{26}, 045501 (2016).





 \end{thebibliography}
\end{document}